\documentclass[twocolumn,showpacs,pre,aps,superscriptaddress]{revtex4}
\usepackage{graphicx}
\usepackage{amssymb}
\usepackage{amsmath}
\usepackage{color}

\begin{document}

\title{Relevance of the eigenstate thermalization hypothesis for thermal relaxation}

\author{Abdellah Khodja}%
\email{akhodja@uos.de}%
\affiliation{Department of Physics, University of Osnabr\"uck, D-49069 Osnabr\"uck, Germany}%

\author{Robin Steinigeweg}%
\email{r.steinigeweg@tu-bs.de}%
\affiliation{Institute for Theoretical Physics, Technical University Braunschweig, D-38106 Braunschweig, Germany}%

\author{Jochen Gemmer}%
\email{jgemmer@uos.de}%
\affiliation{Department of Physics, University of Osnabr\"uck, D-49069 Osnabr\"uck, Germany}%

\begin{abstract}
We study the validity of the eigenstate
thermalization hypothesis (ETH) and its role for the occurrence
of initial-state independent (ISI) equilibration in closed
quantum many-body systems. Using the concept of dynamical
typicality, we present an extensive numerical analysis of energy
exchange in integrable and nonintegrable spin-$1/2$ systems of large
size outside the range of exact diagonalization. In case of
nonintegrable systems, our finite-size scaling shows that the ETH becomes
valid in the thermodynamic limit and can serve as the underlying mechanism
for ISI equilibration. In case of integrable systems,  however, indication 
of ISI equilibration has been observed despite the violation of the ETH. We establish
a connection between this observation and the need of choosing a proper
parameter within the ETH.
\end{abstract}

\date{\today}


\maketitle

\section{Introduction}

Due to experiments in ultracold atomic gases
\cite{trotzky2008, hofferberth2007, bloch2008, cheneau2012,
langen2013}, the question of thermalization in closed quantum
systems has experienced an upsurge of interest in recent years. It
is, however, a paradigm of standard thermodynamical processes that
the final equilibrium state is generally independent of the details
of the initial state. But this general initial-state independence
(ISI) is challenging to proof from an underlying theory such as
quantum mechanics. While for local density matrices of subsystems
the issue of necessary and sufficient conditions for ISI is subtle
\cite{lychkovskiy2010, linden2010}, for expectation values of
observables a widely accepted {\it sufficient} condition for ISI is
the validity of the eigenstate thermalization hypothesis (ETH)
\cite{deutsch1991, srednicki1994, rigol2008}. It claims that the
expectation values of a given observable $A$ should be similar for
energy eigenstates $|n\rangle$ if the energy eigenvalues $E_n$ are
close to each other, i.e., $\langle n | A | n \rangle \approx
\langle n' | A | n' \rangle$ if $E_n \approx E_{n'}$. The relation
to ISI equilibration can be seen from considering the dynamics
\begin{equation}
a(t) := \text{Tr} \{\rho(t) \, A \} = \sum \nolimits_{nm} \rho_{nm}
\, A_{mn} \, e^{\imath (E_m - E_n) t}
\end{equation}
with $A_{mn}= \langle m | A | n \rangle$ and averaging over
sufficiently long times. Given that there are no degeneracies
\cite{reimann2008}, this averaging yields
\begin{equation}
\label{diag}
\bar{a} \approx \sum \nolimits_{n} \rho_{nn} \, A_{nn} \, .
\end{equation}
If, as the  ETH claims, all $A_{nn}$ from an energy region around
$E$ are similar, i.e., $A_{nn} \approx A(E)$, then obviously
$\bar{a} \approx A(E)$ regardless of the initial state $\rho(0)$, as
long as it is also restricted to the same energy region. Hence, if
various $a(t)$ from an energy shell equilibrate at all (for
conditions on that see Refs.\ \cite{reimann2008, linden2010}), they
must do so at the same value $A(E)$, i.e., ISI applies.

Much less clear is whether or not the validity of the ETH is also a
{\it  physically necessary} condition for ISI in the sense that no
{\it relevant} set of states can exhibit ISI without the ETH being
fulfilled. (It is surely necessary if one requires that all states
exhibit ISI.) Equation (\ref{diag}) allows for mathematically
constructing different initial states with the same $\bar{a}$ even
in cases where the ETH does not apply (see, e.g., Refs.\
\cite{delcampo2011, riera2012}). Thus, the ETH may not be a
physically necessary condition. However, the ETH has numerically
been found to be fulfilled for a variety of systems and observables,
and it is commonly expected that the ETH applies to few-body
observables in nonintegrable quantum systems \cite{deutsch1991,
srednicki1994, rigol2009, biroli2010}. But this expectation is
still lacking a rigorous proof. Therefore, the crucial question is:
Is the ETH the ``force that drives'' physical equilibration?

\begin{figure}[t]
\includegraphics[width=0.7\columnwidth]{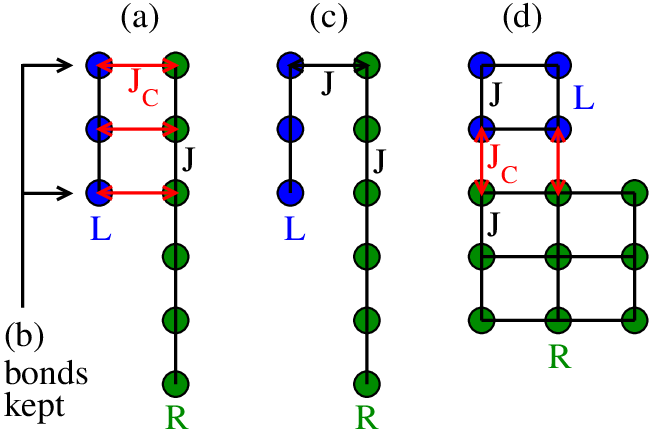}
\caption{(color online) Schematic representation of the systems for
the study of energy exchange between a ``left'' (L) and ``right''
(R) part.} \label{Fig1}
\end{figure}

In the present paper, we consider this question by two different
numerical approaches: i) We study the ETH concerning energy
exchange for a variety of coupled subsystems. From the second
law, ISI equilibration is strongly expected for each setting.
ii) We investigate equilibration dynamics in systems for which
the ETH is clearly violated. In both approaches, we apply the
concept of dynamical typicality \cite{goldstein2006, bartsch2009,
elsayed2013, steinigeweg2014-1, steinigeweg2014-2} and provide a
careful finite-size scaling since, for any finite system, the ETH
is never strictly fulfilled (i.e., $A_{nn} \neq A(E)$ except for
trivial cases) \cite{ikeda2013, steinigeweg2013, beugeling2014,
steinigeweg2014-1}. Our study in i) unveils that the ETH is
fulfilled for energy exchange in nonintegrable systems and
particularly approaches the thermodynamic limit according to a
power-law dependence on the effective Hilbert-space dimension.
Our investigations in ii) are different from other approaches
based on quantum quenches \cite{rigol2006, rigol2007, rigol2009,
santos2012, he2013, gudyma2013} or a period of time-dependent
driving \cite{monnai2011}. Still, we use pure initial states. But
we prepare initial states with the property that the observable of
interest deviates largely from its equilibrated value, while these
states are still restricted to an energy shell. We consider this
subclass of all possible initial states to be generic for
equilibration experiments. As a main result, we observe ISI
relaxation of energy exchange in an integrable system where the
ETH clearly breaks down. In this way, we unveil the need of
choosing a proper ETH-violation parameter.

This paper is structured as follows: In the next
Sec.\ \ref{method} we introduce the numerical method used as well as
the models and observable studied. Then we summarize and discuss in
Sec.\ \ref{nonintegrable} our results on the ETH and ISI
equilibration in nonintegrable systems. The following Sec.\
\ref{nonintegrable} is devoted to the relationship between the ETH
and ISI equilibration for the specific case of integrable systems and
the above mentioned ``observable-displaced'' initial states. In the
last Sec.\ \ref{summary} we summarize and draw conclusions.

\section{Method, Models, and Observable} \label{method}

Convenient parameters to quantify the ETH w.r.t.\ a given
Hamiltonian $H$ and observable $A$ are
\begin{equation}
\bar{A} = \sum_{n=1}^d p_n \, A_{nn} \, , \quad \Sigma^2 =
\sum_{n=1}^d p_n \, A_{nn}^2 - \bar{A}^2 \, , \label{ETH1}
\end{equation}
where $A_{nn}$ are diagonal matrix elements w.r.t.\ the Hamiltonian
eigenstates $| n \rangle$ with eigenvalues $E_n$, $p_n \propto
e^{-(E_n - \bar{E})^2 / 2 \sigma^2}$ is a probability distribution
centered at $\bar{E}$, and $d$ is the Hilbert-space dimension. The
quantities $\bar{A}(\bar{E},\sigma)$ and $\Sigma(\bar{E},\sigma)$
are obviously functions of $\bar{E}$ and an energy width $\sigma$.
Routinely, the ETH is said to be fulfilled if $\Sigma$ is small.
However, whenever $\sigma$ is finite, $\Sigma$ can only be zero for
vanishing slopes of $\bar{A}$, i.e., $\partial \bar{A}/\partial
\bar{E} = 0$. Therefore, for finite $\sigma$, the ETH is fulfilled
if $\Sigma \rightarrow \partial \bar{A} / \partial \bar{E} \,
\sigma$ in the limit of large system sizes \cite{steinigeweg2014-1}.

\begin{figure}[t]
\includegraphics[width=0.8\columnwidth]{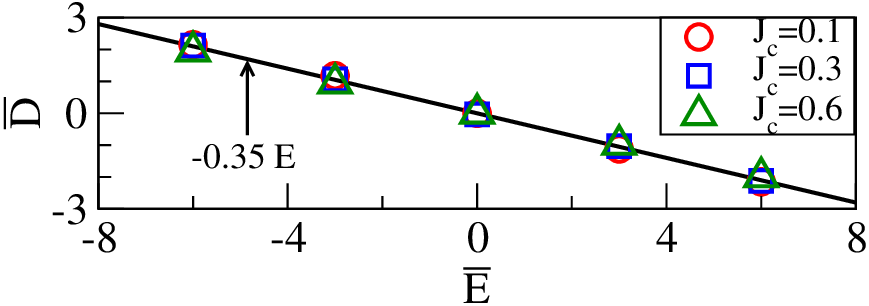}
\caption{(color online) Mean energy difference $\bar{D}$ of energy
eigenstates in an energy shell of width $\sigma = 0.6$ around
$\bar{E}$, calculated for the model in Fig.\ \ref{Fig1} (a) with
anisotropy $\Delta = 0.3$ and system size $N_\text{L} = 8$. For the
``average eigenstate'', the distribution of energy onto the
subsystems is proportional to their sizes.} \label{Fig2}
\end{figure}

A crucial point in this paper is of course finite-size scaling. To
render this scaling as convincing as possible, one needs data on
systems as large as possible. Usually, checking the ETH requires
exact diagonalization \cite{steinigeweg2013, beugeling2014} that
is limited to rather small system sizes. Thus, we employ a recently
suggested method \cite{steinigeweg2014-1} that is based on dynamical
typicality and allows for the extraction of information on the ETH
from the temporal propagation of pure states. This propagation can be
performed by iterative algorithms such as Runge-Kutta \cite{elsayed2013,
steinigeweg2014-1, steinigeweg2014-2}, Chebyshev \cite{deraedt2006,
deraedt2007}, etc.\ and is feasible for larger system sizes. We
use a fourth-order Runge-Kutta iterator with a sufficiently small
time step. Due to typicality-related reasons, the so-computed
quantities $\bar{A}$ and $\Sigma$ are subject to statistical errors.
These errors, however, turn out to be smaller than the symbol sizes
used in this paper \cite{steinigeweg2014-2}.

We study systems consisting of two weakly coupled XXZ spin-$1/2$
chains (except for one example discussed below). The Hamiltonian of
the chains has open boundary conditions and is given by
\begin{eqnarray}
H_j &=& J \sum_{i=1}^{N_j-1} (S_{i, j}^x S_{i+1, j}^x + S_{i,
j}^y S_{i+1, j}^y + \Delta \, S_{i, j}^z S_{i+1, j}^z) \nonumber \\
&+& \sum_i^{N_j} h_{i,j} \,  S_{i,j}^z \, , \label{H1}
\end{eqnarray}
where $j = \text{L}$, $\text{R}$ labels the ``left'' and ``right''
chain. $N_\text{L}$, $N_\text{R}$ are the respective numbers of
spins in the chains, see Figs.\ \ref{Fig1} (a)-(c). $J = 1$ is the
antiferromagnetic exchange coupling constant and $\Delta$ is the
exchange anisotropy. $h_{i,j}$ are spatially random magnetic fields
in $z$ direction that are drawn at random according to a uniform
distribution between $-W/2$ and $W/2$. This way of introducing
disorder is similar to the random on-site potential of the Anderson
model, widely investigated in the context of many-body localization
\cite{huse2010}.

To allow for energy exchange, we add a coupling term $H_\text{C}$
between the two chains: $H = H_\text{L} + H_\text{R} + H_\text{C}$.
This term has a very similar form and reads
\begin{equation}
H_\text{C} = J_\text{C} \sum_{i=1}^{N_\text{L}} c_i \, (S_{i,
\text{L}}^x S_{i, \text{R}}^x + S_{i, \text{L}}^x S_{i, \text{R}}^x
+ \Delta \, S_{i, \text{L}}^z S_{i, \text{R}}^z) \, , \label{H2}
\end{equation}
where $J_\text{C}$ is the coupling strength. For $c_i = 1$, the
total Hamiltonian $H$ represents a nonintegrable structure of ladder
type while, for $c_i = \delta(i,1)$, $H$ reduces to a single chain.
This chain is integrable for $J = J_\text{C}$ and $W=0$ only. We
also consider the intermediate case with $c_i = \delta(i,1) +
\delta(i,N_L)$ and a two-dimensional situation with $c_i = 1$: In
this situation the chains in Eq.\ (\ref{H1}) are replaced by $N_j
\times N_j$ lattices with interactions between all nearest neighbors
and the coupling in Eq.\ (\ref{H2}) is replaced by an interaction at
the contact of the lattices, as illustrated in Fig.\ \ref{Fig1} (d).

The observable $A$ we are going to investigate for accord with the
ETH is the energy difference. $A$ is represented by the operator $D
= H_L - H_R $. For any left-right symmetric model and observable,
the ETH is necessarily fulfilled. Thus, to avoid this trivial case,
we choose right chains to consist of twice as much spins as the left
chains throughout this paper, see Figs.\ \ref{Fig1} (a)-(c). In the
two-dimensional situation we choose the right-lattice side to have
one spin more than the left-lattice side.

\section{Results on Nonintegrable Systems} \label{nonintegrable}

First, we consider $\bar{D}(\bar{E})$, i.e., the mean energy
difference for the energy eigenstates in a small energy interval
centered at $\bar{E}$. We control the smallness of this energy
interval by choosing $\sigma = 0.6$. (This choice is kept for the
remainder of this paper.) Results on the model in Fig.\ \ref{Fig1}
(a) are displayed in Fig.\ \ref{Fig2} for various coupling strengths
$J_\text{C}$ at fixed anisotropy $\Delta = 0.3$ and system size $N_L
= 8$.

To gain insight into the results in Fig.\ \ref{Fig2}, let us
consider the following analog to the standard equipartition theorem:
Assume that every term in the Hamiltonian (``bond'') contains for
energy eigenstates an amount of energy proportional to its strength.
For weak coupling, the energies corresponding to terms of
$H_\text{C}$ can be neglected. Then one expects a left/right-chain
partition of energy as $(N_L-1)/(N_R-1) = 7/15$, yielding an energy
difference $\bar{D} = (N_L - N_R)/(N_R + N_L -2) \, \bar{E} =- 4
\bar{E}/11$. This expectation is indicated by the solid line in
Fig.\ \ref{Fig2}. Clearly, the equipartition assumption is well
justified for a wide range of interactions strengths $J_\text{C}
\leq 0.6$. We find this principle to hold for each model addressed
in this paper, even though not shown explicitly here. This finding
is a first main result, although not at the center of the
investigation in this paper.

\begin{figure}[t]
\includegraphics[width=1.0\columnwidth]{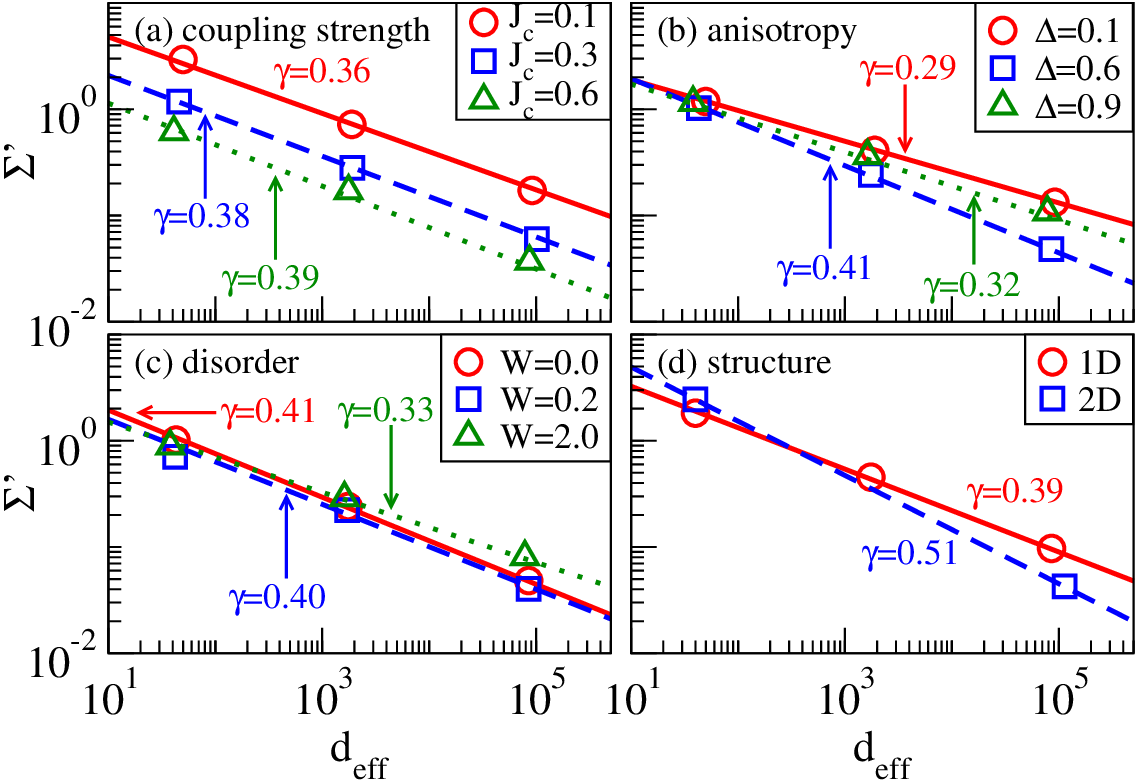}
\caption{(color online) The parameter $\Sigma'$, quantifying the
degree of ETH violation, as a function of the effective dimension
$d_\text{eff}$ that increases with system sizes. The violation
appears to vanish according to a power law in the limit of large
system sizes, regardless of the properties of the specific models
studied in (a)-(d). For model parameters, see text.} \label{Fig3}
\end{figure}

Next we turn to the ETH. To limit computational effort, we focus on
the energy regime around $\bar{E} = 0$, however, we have found
similar results for other points in energy space. For the $E=0$
regime we compute both, $d_\text{eff} = \text{Tr} \{ e^{-(
H-\bar{E})^2/2 \sigma^2} \}$ and $\Sigma' = \Sigma - \partial
\bar{D} / \partial \bar{E} \, \sigma$. $d_\text{eff}$ has the
meaning of the number of eigenstates that constitute the
``relevant'' energy shell. $\Sigma'$ is the quantity that is
expected to approach zero if the ETH is fulfilled. (We get $\partial
\bar{D}/\partial \bar{E} \, \sigma = 0.21$ throughout this paper.)
For all models studied, we show in Fig.\ \ref{Fig3} a double
logarithmic plot of $\Sigma'$ as a function of $d_\text{eff}$ that
increases with system size.

In Figs.\ \ref{Fig3} (a)-(c) we summarize our results on $\Sigma'$
for the model in Fig.\ \ref{Fig1} (a): In Fig.\ \ref{Fig3} (a) we
vary the coupling strength $J_\text{C}$ at fixed anisotropy $\Delta
= 0.3$; In Fig.\ \ref{Fig3} (b) we vary the anisotropy $\Delta$ at
fixed coupling strength $J_\text{C} = 0.3$; and in Fig.\ \ref{Fig3}
(c) we study cases of weak and intermediate amount of disorder for
$\Delta = 0.6$ and $J_\text{C} = 0.3$. In Fig.\ \ref{Fig3} (d) we
show results for the same values $\Delta = 0.6$ and $J_\text{C} =
0.3$ but for $W = 0$ and the different coupling structure in Fig.\
\ref{Fig1} (b) and the two-dimensional case in Fig.\ \ref{Fig1} (d).

In all cases, $\Sigma'$ apparently scales with $d_\text{eff}$ as a
power law $\Sigma' \propto d_\text{eff}^{-\gamma}$, which is in
accord with the findings in Ref.\ \cite{beugeling2014}. Consequently,
the ETH is fulfilled in the limit of large system sizes. This is our
second main result. It indeed indicates that the relaxation of the
energy difference between weakly coupled quantum objects may be
generically expected for all sorts of initial states. For a
random-matrix model, one would expect a power-law scaling with the
exponent $\gamma = 0.5$. For all nonintegrable cases we studied we
find a slightly different exponent. If we call closeness to $\gamma
= 0.5$ the ``amount'' of nonintegrability, then the two-dimensional
case in Fig.\ \ref{Fig1} (d) is the most nonintegrable one studied
here. For the one-dimensional cases, the amount of nonintegrability
does not depend crucially on the coupling constant $J_\text{C}$ in
the small $J_\text{C}$ regime, while the dependence on the
anisotropy $\Delta$ is nonmonotonic.

\section{Microcanonical Observable, Displaced States and Results on Integrable Systems}

\begin{figure}[t]
\includegraphics[width=0.8\columnwidth]{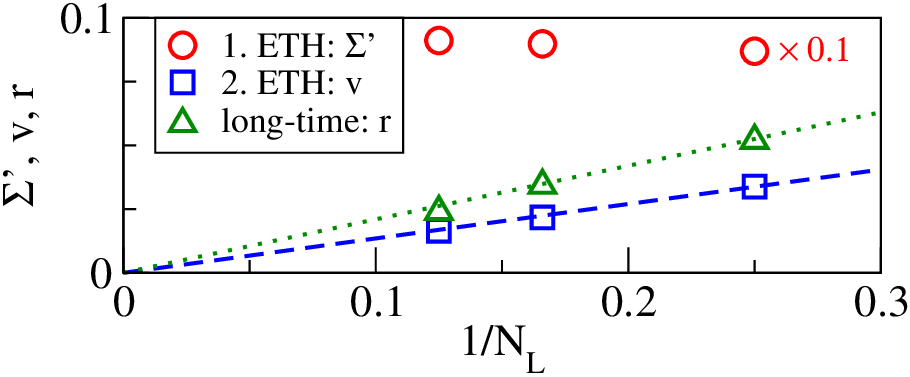}
\caption{(color online) The ETH parameter $\Sigma'$ (circles), the
scaled ETH parameter $v$ (squares), and the long-time value $r$ for
the integrable Heisenberg chain in an energy window of width $\sigma
= 0.6$ around energy $\bar{E}=0$. All quantities are shown as a
function of the inverse number of spins $1/N_\text{L}$.} \label{Fig4}
\end{figure}

The above findings certainly raise immediately the question of ISI
equilibration in integrable models: Is it reasonable to expect the
violation of ISI equilibration for two subsystems which are
connected to form an integrable system? Comparable investigations
found answers in both directions \cite{kinoshita2006, rigol2007,
santos2012}, even for nonintegrable systems
\cite{gogolin2011,rossini,rigol2012}. To analyze this question, we
study a pure Heisenberg chain, from the point of view depicted in
Fig.\ \ref{Fig1} (c). As already mentioned, this Heisenberg chain is
integrable for $J_\text{C} = 1$,  $\Delta=0.6$ and $W = 0$. For this
model, we show in Fig.\ \ref{Fig4} (circles) results on $\Sigma'$ as
a function of $1/N_\text{L}$.

Unlike all previous examples, $\Sigma'$ does not decrease as system
size increases, which is in accord with the system being integrable.
To clarify the impact on relaxation, we analyze the dynamics of
``microcanonical'' and ``observable-displaced'' initial states
$\rho_\text{MOD}$,
\begin{equation}
\label{mod}
 \rho_\text{MOD}:\propto e^{-(H^2 + \beta^2 [D-d(0)]^2)/ 2 \sigma^2}
\, .
\end{equation}
Obviously, the parameter $d(0)$ controls the degree of observable
displacement. The energy is concentrated within a width $\sigma$
around zero. Since $H$ and $D$ do not commute, we need to find a
compromise between energy concentration and observable displacement,
done by tuning $\beta$. We choose $\sigma=0.6$, $\beta = 0.5$, and
$d(0) = \pm N_L$. This state may be viewed as being based on Jayne's
principle: It represents the maximum-entropy state under given means
and variances for energy and observable.\\ A comment
on the usage of this type of initial states as opposed to initial
states generated by quantum quenches may be instructive. While
quantum-quench approaches produce a complex, non-stationary state,
it may or may not make the considered observable deviating from its
long-time average. The MOD state, on the other hand, is deliberately
designed to obtain a specific initial value of the considered
observable while keeping the energy reasonably well-defined. Thus,
the MOD approach facilitates the generation of initial values of the
observable that are ``substantially off-equilibrium'', to put it in a
catchy phrase. Using the above parameters, we indeed get $\text{Tr}
\{ \rho_\text{MOD} \, D(0) \} \approx d(0) = \pm N_L$. The largest
eigenvalue of $D$ is upper-bounded by $D \leq 9/4 (N_L-1)$. Hence,
the initial expectation value of $D$ reaches at least $50\%$ of the
difference between its highest possible value and its long-time
average. Corresponding values for standard quench dynamics are
usually much lower \cite{rigol2012, rigol2009}. \\ Note also that such a
MOD state does not necessarily feature a smooth probability
distribution w.r.t.\ energy, as suggested in Ref.\ \cite{ikeda2011}
to explain ISI.

Numerically, $\rho_\text{MOD}$ is challenging to compute for systems
beyond the reach of exact diagonalization. Thus, we prepare the
partially random state
\begin{equation}
| \phi_\text{MOD} \rangle = \langle \varphi | \rho_\text{MOD} |
\varphi \rangle^{-1/2} \, \rho_\text{MOD}^{1/2} \, | \varphi \rangle
\, , \label{init}
\end{equation}
where $|\varphi \rangle$ is a random state drawn according to the
unitary invariant (Haar-) measure. In the context of typicality,
e.g, from Ref.\ \cite{bartsch2009, steinigeweg2014-2}, it may be
inferred that the dynamics of $| \phi_\text{MOD} \rangle$ are
similar to those of $\rho_\text{MOD}$ w.r.t.\ to the observable,
i.e.,
\begin{equation}
d(t) = \langle \phi_\text{MOD}| D(t) | \phi_\text{MOD} \rangle =
\text{Tr} \{\rho_\text{MOD} \, D(t)\} + \epsilon \, ,
\end{equation}
where $\epsilon$ is a random variable with zero mean. The standard
deviation of $\epsilon$ is upper bounded by $(\text{Tr} \{
\rho_\text{MOD} \, D^4\}/d_\text{eff})^{1/2}$. Thus, since the
density of states is large at $\bar{E}=0$, we get $\epsilon \approx
0$.

\begin{figure}[t]
\includegraphics[width=0.8\columnwidth]{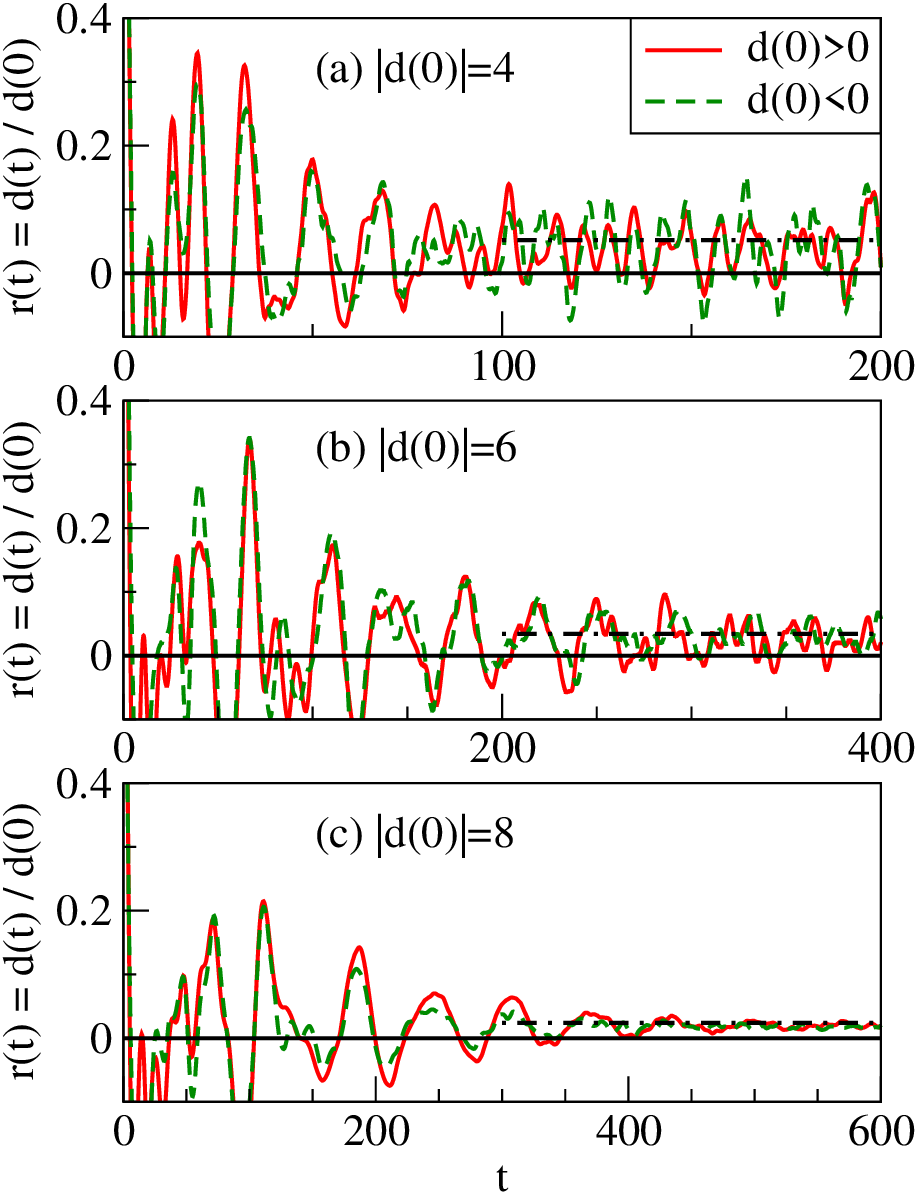}
\caption{Real-time decay of energy-difference expectation values $d(t)$,
divided by their initial values $d(0)$, for initial states in Eq.\
(\ref{init}) and three system sizes: (a) $N_\text{L} = 4$, (b)
$N_\text{L} = 6$, and (c)  $N_\text{L} = 8$. While $r(t) = d(t) / d(0)$
does not vanish for long times and finite systems, comparing (a), (b),
and (c) indicates a vanishing $r(t)$ for long times in the limit of
large system sizes. For more details on finite-size scaling, see Fig.\
\ref{Fig4} (triangles).} \label{Fig5}
\end{figure}

Using Runge-Kutta we can prepare these initial states and compute
the time evolution of energy-difference expectations $d(t)$ divided
by their initial values $d(0)$, i.e., $r(t) = d(t)/d(0)$. In Fig.\
\ref{Fig5} we depict our results for three system sizes $N_\text{L}
=4$, $6$ and $8$. Remarkably, the scaled differences $r(t)$ are
practically independent of the sign of the initial value $d(0)$.
While the energy difference $r(t)$ decays, it does not decay all the
way to zero. In fact, the fluctuations around  the corresponding nonzero 
equilibrium value decrease for increasing system size $N_\text{L}$. Thus,
for finite systems, there is indeed no clean ISI equilibration. Since
we find the same behavior for the overwhelming majority of all states
prepared according to Eq.\ (\ref{init}), it is reasonable to claim
that the validity of the ETH is imperative for ISI equilibration.
However, comparing Figs.\ \ref{Fig5} (a), (b), and (c) indicates that
$r(t)$ decreases at large times as system size increases.
(Considering $r(t)$ at large times as a central quantity is also
suggested in Ref.\ \cite{yurovsky2011}.) This
finding implies that ISI equilibration may be expected for this
specific system and observable in the limit of large system sizes,
regardless of the ETH being fulfilled. While it is well-known that
also integrable systems may exhibit ISI (see the corresponding
statement at the beginning of this section), these examples either
refer to situations where the ETH may apply w.r.t.\ the considered
observable regardless of integrability \cite{rossini} and/or initial
states generated from quenches that are not tailored to make the
considered observable initially deviating largely from its long-time
average \cite{rigol2012}. In contrast to that, our finding addresses
the occurrence of ISI for a situation where the violation of the ETH
is numerically evident and the initial state is specifically chosen
to initially deviate largely from its long-time average. This is a
third main result of our paper.

The above results naturally lead to the question why larger systems
exhibit equilibration that is closer to ISI even though $\Sigma'$ is
essentially the same for large and small systems. To clarify this
question, let us consider a quantity that takes the ``natural''
scale of the operator into account, rather than just the bare
$\Sigma'$. Therefore, we define a scaled ETH parameter $v$ as $v :=
(\Sigma')^2 / \delta^2$, where $\delta$ is the variance of the
operator spectrum within the respective energy regime, i.e.,
$\delta^2 = \overline{D^2} - \bar{D}^2 \label{delta}$. The bar in
$\overline{D^2}$ and $\overline{D}$ is still defined according to
Eq.\ (\ref{ETH1}). Obviously, even if $\Sigma'$ does not decrease as
system size increases, $v$ may still do so. To find out whether or
not it actually does, one has to compute the two quantities
$\overline{D^2}$ and $\overline{D}$. This computation can be done by
the same typicality-based method used so far in this paper. In
Fig.\ \ref{Fig4} (squares) we show the corresponding result.
Apparently, $v$ vanishes in the limit of large system sizes even
though $\Sigma'$ does not. This observation suggests that $v$ is a
reliable predictor of ISI equilibration that can vanish even if
systems are integrable or close to integrability.

\section{Summary and Conclusion} \label{summary}

In this paper we studied the validity of the ETH and its role for
the occurrence of ISI equilibration in closed quantum many-body
systems. Using the concept of dynamical typicality, we presented an
extensive numerical analysis of energy exchange in integrable and
nonintegrable spin-$1/2$ systems of large size outside the range of
exact diagonalization. In case of nonintegrable systems, our
finite-size scaling showed that the ETH becomes valid in the
thermodynamic limit and can serve as the underlying mechanism for
ISI equilibration. In case of integrable systems, however, indication
of ISI equilibration has been observed despite the violation of
the ETH and initial states that specifically facilitate the
deviation of the observable from its equilibrium value. We
established a connection between this observation and the need of
choosing a proper parameter within the ETH.

\end{document}